\documentclass[twocolumn]{aastex62}

\usepackage{amsmath}
\usepackage[normalem]{ulem}

\graphicspath{{./}{figures/}}

\received{24 June 2020}
\revised{15 September 2020}
\accepted{5 January 2021}
\submitjournal{ApJ}

\shorttitle{CN Zeeman AS 209}
\shortauthors{Harrison et al.}

\begin{document}

\title{ALMA CN Zeeman Observations of AS 209: Limits on Magnetic Field Strength and Magnetically Driven Accretion Rate}

\author[0000-0003-2118-4999]{Rachel E. Harrison}
\affiliation{Department of Astronomy, University of Illinois, Urbana, IL 61801, USA}
\author[0000-0002-4540-6587]{Leslie W. Looney}
\affiliation{Department of Astronomy, University of Illinois, Urbana, IL 61801, USA}
\author[0000-0003-3017-4418]{Ian W. Stephens}
\affiliation{Harvard-Smithsonian Center for Astrophysics, Cambridge, MA 02138, USA}
\affiliation{Department of Earth, Environment and Physics, Worcester State University, Worcester, MA 01602, USA}
\author{Zhi-Yun Li}
\affiliation{Department of Astronomy, University of Virginia, Charlottesville, VA 22903, USA}
\author[0000-0003-1534-5186]{Richard Teague}
\affiliation{Department of Astronomy, University of Michigan, Ann Arbor, MI 48109, USA}
\author{Richard M. Crutcher}
\affiliation{Department of Astronomy, University of Illinois, Urbana, IL 61801, USA}
\author[0000-0002-8537-6669]{Haifeng Yang}
\affiliation{Institute for Advanced Study, Tsinghua University, Beijing, 100084, China}
\author{Erin Cox}
\affiliation{Center for Interdisciplinary Exploration and Research in Astrophysics (CIERA) and Department of Physics andAstronomy, Northwestern University, Evanston, IL 60208, USA}
\author[0000-0001-5811-0454]{Manuel Fern{\'a}ndez-L{\'o}pez}
\affiliation{Instituto Argentino de Radioastronom{\'i}a (CCT-La Plata, CONICET; CICPBA),Buenos Aires, Argentina}
\author[0000-0001-9407-6775]{Hiroko Shinnaga}
\affiliation{Department of Physics and Astronomy, Graduate School of Science and Engineering, Kagoshima University, Amanogawa Galaxy Astronomy Research Center (AGARC), Kagoshima 890-0065, Japan}

\correspondingauthor{Rachel Harrison}
\email{reh3@illinois.edu}




\begin{abstract}
While magnetic fields likely play an important role in driving the evolution of protoplanetary disks through angular momentum transport, observational evidence of magnetic fields has only been found in a small number of disks. Although dust continuum linear polarization has been detected in an increasing number of disks, its pattern is more consistent with that from dust scattering than from magnetically aligned grains in the vast majority of cases. Continuum linear polarization \added{from dust grains aligned to a magnetic field} can reveal information about the magnetic field's direction, but not its strength. On the other hand, observations of circular polarization in molecular lines produced by Zeeman splitting offer a direct measure of the line-of-sight magnetic field strength in disks. We present upper limits on the net toroidal and vertical magnetic field strengths in the protoplanetary disk AS 209 derived from Zeeman splitting observations of the CN 2-1 line. \replaced{The 3$\sigma$ upper limit on the net line-of-sight magnetic field strength in AS 209 is 4.8 mG on the redshifted side of the disk and 4.2 mG on the blueshifted side of the disk. Given the disk's inclination angle, we set a 3$\sigma$ upper limit on the net toroidal magnetic field strength of 8.4 and 8.0 mG for the red and blue sides of the disk, respectively, and 5.9 and 5.6 mG on the net vertical magnetic field on the red and blue sides of the disk. Magnetic fields of this strength would be too weak to drive accretion rates at the $\sim$50 au scale as high as those inferred close to the central protostar. A mismatch between the inferred mass accretion rates near the protostar and on the 10's of au scale, if confirmed, could lead to the rapid depletion of material in the inner disk.}{The 3$\sigma$ upper limit on the net line-of-sight magnetic field strength in AS 209 is 5.0 mG on the redshifted side of the disk and 4.2 mG on the blueshifted side of the disk. Given the disk's inclination angle, we set a 3$\sigma$ upper limit on the net toroidal magnetic field strength of 8.7 and 7.3 mG for the red and blue sides of the disk, respectively, and 6.2 and 5.2 mG on the net vertical magnetic field on the red and blue sides of the disk. If magnetic disk winds are a significant mechanism of angular momentum transport in the disk, magnetic fields of a strength close to the upper limits would be sufficient to drive accretion at the rate previously inferred for regions near the protostar.}

\end{abstract}



\section{Introduction} \label{sec:intro}

Magnetic fields are thought to play an essential role in the evolution of protoplanetary disks by providing a means of angular momentum transport through the magnetorotational instability (MRI) \citep{balbus} or magnetic disk winds \citep{blandford}. The MRI mechanism requires that material have a high enough ionization fraction to be well-coupled to the magnetic field. In protoplanetary disks, the ionization fraction is only expected to be high enough to produce magnetorotational instability within \replaced{ a few tens of AU}{$\sim$0.1 au} of the central protostar, where the temperature is high enough to produce thermal ionization, and in the surface layers of the disk, where non-thermal ionization sources such as cosmic rays, \added{X-rays \citep{1999ApJ...518..848I}, and FUV photons \citep{2011ApJ...735....8P} can penetrate \citep{gammie}.} The poloidal component of the disk magnetic field gives rise to the magnetic disk wind. The disk wind system modeled in \citet{suriano18} predicts that the disk magnetic field will have both toroidal and poloidal components, with the toroidal component being stronger than the poloidal component in regions above and below the disk midplane (see \citealt{suriano18} Figures 4 and 5).

Observationally determining magnetic field strengths and morphologies in protoplanetary disks to constrain the various angular momentum transport mechanisms has proven difficult. The original motivation behind many millimeter and sub-millimeter continuum polarization observations of protoplanetary disks was to find evidence of dust grain alignment to disk magnetic fields \citep[e.g.,][]{2014Natur.514..597S, 2014ApJ...780L...6R, 2015ApJ...798L...2S}. However, the polarized emission seen in many disks at these wavelengths is better explained by dust scattering of thermal emission \citep[e.g.,][]{2015ApJ...809...78K, 2016MNRAS.456.2794Y, 2017MNRAS.472..373Y}. While possible evidence of grain aligment to disk magnetic fields has been found in the circumbinary disks BHB07-11 \citep{2018A&A...616A..56A} and VLA 1623 \citep{2018ApJ...859..165S, 2018ApJ...861...91H}, and in the disk of HD 142527 \citep{2018ApJ...864...81O}, these observations provide information about the direction and morphology of the magnetic field lines, not the magnetic field strength.

The Zeeman effect offers a direct means of constraining the line-of-sight magnetic field strength without contamination from continuum dust scattering; however, it is possible to produce circular polarization in molecular lines through resonant scattering \citep{2013ApJ...764...24H}. In the presence of a magnetic field, spectral lines split apart in frequency to a degree that depends on the magnetic field strength as $\nu = \nu_0 \pm \frac{eB}{4 \pi m_e c}$ \citep[e.g.,][]{2019FrASS...6...66C}, where $\nu_0$ is the line frequency in the absence of a magnetic field and $B$ is the magnetic field strength. In astronomical sources with intrinsically weak magnetic fields compared to those in, for example, stellar photospheres, the observable of Zeeman splitting is circular polarization of the spectral line, which measures the line of sight magnetic field strength. Paramagnetic species, such as the CN radical, are particularly sensitive to the Zeeman effect. CN is also one of only three species in which the Zeeman effect has been unambiguously detected in extended molecular gas in star forming regions \citep{2012ARA&A..50...29C}. However, the Zeeman effect has yet to be detected \replaced{in the disk plane}{in a protoplanetary disk}, despite a recent attempt with ALMA \citep{2019A&A...624L...7V}. 

AS 209 is a protoplanetary disk located in the Ophiuchus star forming region at a distance of 126 pc from the Sun \citep{2016A&A...595A...2G}. The protostar has a mass of 1.25 $M_\odot$ \citep{2018ApJ...868..113T} and a luminosity of 1.5 $L_\odot$ \citep{2016A&A...588A..53T}. The disk is known to have two rings at $r$ = 75 au and $r$ = 130 au and two gaps at $r$ = 62 au and $r$ = 103 au \citep{2018A&A...610A..24F}. This source was selected as a target source based on its inclination angle, the presence of a bright CN line \citep{2011ApJ...734...98O}, and high accretion rate of nearly $10^{-7}$ $M_\odot$ yr$^{-1}$ \citep{2000ApJ...539..815J}. The disk's inclination angle ($i$) of 35.3$^\circ$ \added{$\pm 0.8^\circ$} \citep{2018A&A...610A..24F}, where $i = 0^\circ$ represents a face-on disk, means that both toroidal and vertical magnetic field lines would have a component along the line of sight, with the toroidal component $B_\phi = B_{LOS}/\sin i$ along the disk's major axis and the vertical component $B_z = B_{LOS}/\cos i$. Finally, a strong magnetic field would likely be needed to drive its high accretion rate.

In this paper, we present Zeeman observations toward AS 209 using CN 2-1.  In Section 2, we present the observations, including the continuum polarization. Without an obvious Zeeman detection, we then use \replaced{various stacking approaches}{two analysis approaches} in Section 3. In Section 4, we place limits on the magnetic strength, and in Section 5, we summarize the results.

\section{Observations}
These observations were taken with the Atacama Large Millimeter/Sub-millimeter Array (ALMA). All data discussed in this paper were taken as part of ALMA project 2018.1.01030.S (PI: Rachel Harrison). The observations were taken in four execution blocs between 6 March and 7 March 2019 while the array was in configuration C43-1. The total observing time was 6.1 hours, of which 2.7 hours were spent on AS 209. The average sampling time was $\sim$6 seconds. J1733-1304 was the phase calibrator, J1751+0939 was the polarization calibrator, and J1427-4206 was the bandpass calibrator. The dataset consists of two spectral line windows with a channel width of 122.070 kHz and a total bandwidth of 117.1875 MHz each, and two continuum windows with a channel width of 976.562 kHz and a total bandwidth of 937.5000 MHz each. The spectral line windows are centered on  226.64013 GHz and 226.88081 GHz. One spectral line window covered lines 1-4 listed in Table 1, and the other covered lines 5-9. \replaced{The Zeeman factor for line 1 is from \citet{2019A&A...624L...7V}, and the factors for all of the other lines are from Shinnaga and Yamamoto (in preparation).}{The Zeeman factors for all of the lines observed are from Shinnaga and Yamamoto (in preparation).} Shinnaga and Yamamoto calculated \added{the Zeeman factors} under the framework of the first-order perturbation, as the interstellar magnetic field is quite weak, as weak as 100 $\mu$G or less. The authors employ the Hund’s case (b) for the coupling scheme of the angular momenta for the calculation \added{\citep{gordy1970}}. We quote Shinnaga and Yamamoto's Zeeman factor values to the second decimal place, and they are in agreement with those reported in \citet{2019A&A...624L...7V}.

\begin{center}
{\bf Table 1: CN N=$2 \rightarrow 1$ Hyperfine Lines}
\vskip 0.01in
\begin{tabular}{cccc}
\hline\hline
Line & $J,F \rightarrow J^\prime,F^\prime$ & $\nu$ (GHz) & Z (Hz/$\mu$G) \\
\hline
1 & $\frac{3}{2},\frac{3}{2} \rightarrow \frac{1}{2},\frac{3}{2}$ & 226.63217 & $-0.72$ \\ [1mm]
2 & $\frac{3}{2},\frac{5}{2} \rightarrow \frac{1}{2},\frac{3}{2}$ & 226.65956 & $-0.71$ \\ [1mm]
3 & $\frac{3}{2},\frac{1}{2} \rightarrow \frac{1}{2},\frac{1}{2}$ & 226.66369 & $-0.62$ \\ [1mm]
4 & $\frac{3}{2},\frac{3}{2} \rightarrow \frac{1}{2},\frac{1}{2}$ & 226.67931 & $-1.18$ \\ [1mm]
5 & $\frac{5}{2},\frac{5}{2} \rightarrow \frac{3}{2},\frac{3}{2}$ & 226.87419 & +0.71 \\ [1mm]
6 & $\frac{5}{2},\frac{7}{2} \rightarrow \frac{3}{2},\frac{5}{2}$ & 226.87478 & +0.40 \\ [1mm]
7 & $\frac{5}{2},\frac{3}{2} \rightarrow \frac{3}{2},\frac{1}{2}$ & 226.87590 & +1.18 \\ [1mm]
8 & $\frac{5}{2},\frac{3}{2} \rightarrow \frac{3}{2},\frac{3}{2}$ & 226.88742 & +1.47 \\ [1mm]
9 & $\frac{5}{2},\frac{5}{2} \rightarrow \frac{3}{2},\frac{5}{2}$ & 226.89213 & +1.06 \\ [1mm]
\hline
\end{tabular}
{\replaced{Frequencies and Zeeman splitting factors (Z) calculated from theory for the hyperfine lines (Shinnaga and Yamamoto in preparation and \citealt{2019A&A...624L...7V}) covered in these observations.}{Frequencies and Zeeman splitting factors (Z) calculated from theory for the hyperfine lines (Shinnaga and Yamamoto in preparation) covered in these observations.} The Zeeman splitting factor is a measure of how much a line will be split in frequency by a given magnetic field strength. A higher absolute value of $Z$ means that a line is more sensitive to the line of sight magnetic field.}
\end{center}

\replaced{The data were calibrated at the North American ALMA Science Center using the standard polarization calibration procedure. Because these observations were taken in a non-standard mode, the data were calibrated by hand at the NAASC rather than with the ALMA science pipeline.}{The data were calibrated by data analysts at the NAASC using a script developed for calibrating ALMA polarization observations.} All data reduction was performed using the Common Astronomy Software Applications (CASA) version 5.4.0. The data were cleaned using Briggs weighting with a robust parameter of 0.5 to create image cubes and continuum images for all four Stokes parameters. We performed one round of phase-only self calibration on the continuum $I$ data, with the solution interval set to the scan length. Before making the line image cubes, we subtracted the continuum emission from the spectral line windows. The image cubes were created with a spectral resolution of 0.25 km/s. The images have a beam size of 1.40$\arcsec$ $\times$ 1.27$\arcsec$. The linear polarized intensity map was debiased using the average noise value determined from the $Q$ and $U$ maps, an estimator used by e.g. \citet{1974ApJ...194..249W} and \citet{2016MNRAS.461..698V}: 

\begin{equation}
  P=\begin{cases}
    \sqrt{Q^2 + U^2 - \sigma^2} & \text{if $\sqrt{Q^2 + U^2} \geq \sigma$} \\
    0 & \text{otherwise}
  \end{cases}
\label{eq:debias}
\end{equation}

We estimate that the uncertainty on the amplitude calibration is $\pm$10\%, and from here on, we only give statistical uncertainties. The rms values of the image cubes are $\sigma_{(I, Q, U, V)}$ = (0.91, 0.95, 0.96, 0.95) mJy beam$^{-1}$ per 0.25 km/s channel in the spectral window containing lines 1-4 and $\sigma_{(I, Q, U, V)}$ = (0.95, 0.88, 0.90, 0.89) mJy beam$^{-1}$ per 0.25 km/s channel in the spectral window containing lines 5-9. The continuum I rms value was 0.60 mJy beam$^{-1}$ before self calibration and 0.33 mJy beam$^{-1}$ after self calibration. The rms values for the $Q$, $U$, and $V$ continuum images are $\sigma_{( Q, U, V)}$ = (0.015, 0.012, 0.013) mJy beam$^{-1}$. The higher noise value in Stokes $I$ compared to $Q$, $U$, and $V$ is due to dynamic range limits. 

\section{Results}
In the dust continuum total intensity (Stokes $I$), the disk has a peak flux of 127.4 mJy/beam and a total intensity of 215.0 $\pm$  1.8 mJy.
All nine of the hyperfine components that we targeted were detected in Stokes $I$. The integrated line intensity for the hyperfine components of lines 5-7 in Table 1 is 3.874 $\pm$ 0.015 Jy km/s, which is consistent with the value of 3.32 $\pm$ 0.14 Jy km/s from \citep{2011ApJ...734...98O}, given that the absolute flux calibration uncertainty is $\sim$10\% for ALMA and $\sim$10-15\% for the SMA.
The velocity map of the source shows the pattern expected from a rotating disk (see Figure 1b). 
\begin{figure*}[ht!]
\gridline{\fig{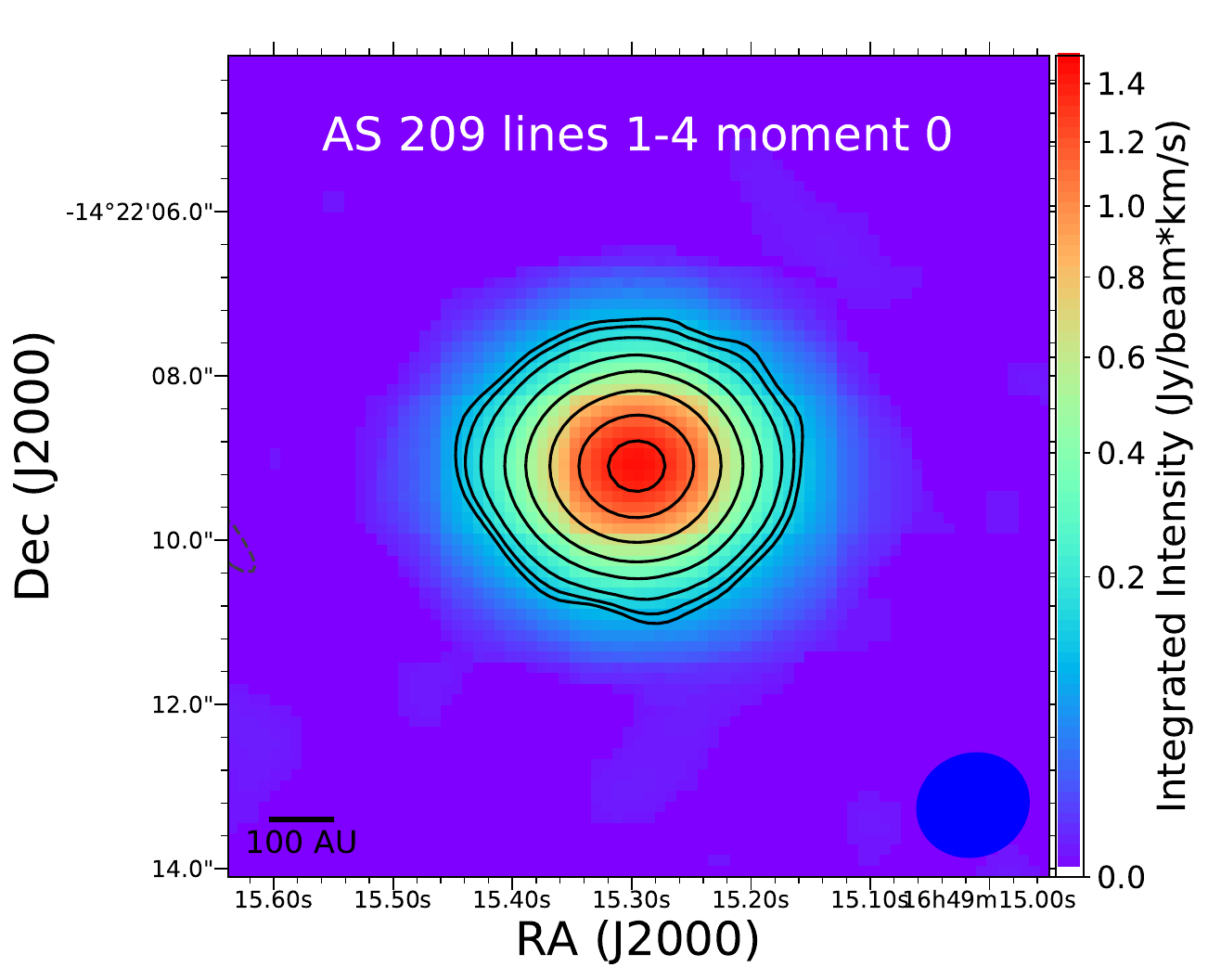}{0.45\textwidth}{(a)}
          \fig{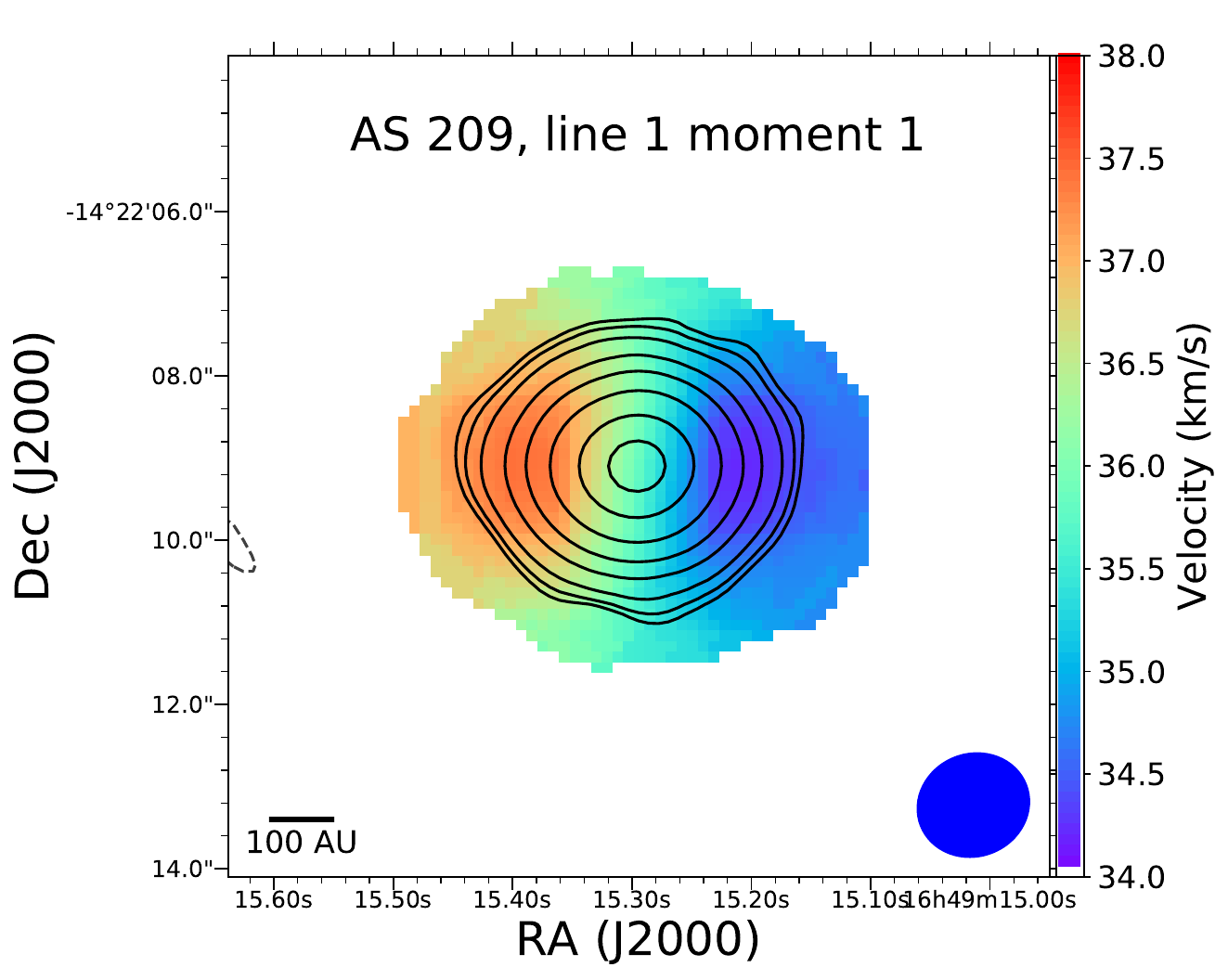}{0.45\textwidth}{(b)}
          }
\caption{Integrated intensity (moment 0) map for hyperfine components 1-4 (a), and velocity (moment 1) map for hyperfine 1 (b). The moment maps are created from line emission with intensity $\geq 5\sigma_{I line}$. Contours represent total intensity (Stokes $I$) of 3, 10, 50, 100, 200, 325 and 500$\sigma$ levels.} 
\label{moment}
\end{figure*}
\subsection{Line circular polarization}
The Stokes $V$ polarization produced by the Zeeman effect is related to the strength of the line of sight B-field $B_{LOS}$, the derivative of the line Stokes $I$ flux with respect to frequency $dI/d\nu$, and the Zeeman splitting factor $Z$ by $V = \frac{1}{2} Z B_{LOS}\frac{dI}{d\nu}$ in the case where the frequency splitting is small compared to the Stokes $I$ line width. 
We made maps of all of the CN hyperfine lines in Table 1 looking for detections in Stokes $V$.  \deleted{Unfortunately,} Stokes $V$ was not obviously \replaced{seen}{detected} in the channel maps, spectra, or moment 0 maps. However, the signal could be contaminated by instrumental terms. To solve for and eliminate instrumental effects that could mimic Zeeman splitting, we use the technique developed by \citet{1996ApJ...456..217C}. This technique fits the observed Stokes $V$ profile of each hyperfine component to the expression
\begin{equation}
V_j(\nu) = C_1 I_j(\nu) + C_2 \frac{dI_j(\nu)}{d\nu} + \frac{1}{2}B Z_j  \frac{dI_j(\nu)}{d\nu}
\label{eq:fit}
\end{equation}
where $j$ refers to each hyperfine component. $C_1$ absorbs any gain difference between left and right polarization that was not calibrated out previously, as well as any linearly polarized line signal. C$_2$ absorbs any instrumental polarization effects such as beam squint that produce pseudo-Zeeman splitting, which shows up as the same splitting in each hyperfine line. \replaced{$C_3$}{$B$} will be non-zero only if there is circular polarization due to the Zeeman effect. We used this method to simultaneously fit all the unblended portions of the lines integrated over the circumstellar disk, but we were still unable to find a detection of Zeeman splitting.

\begin{figure}
    \includegraphics[width=\linewidth]{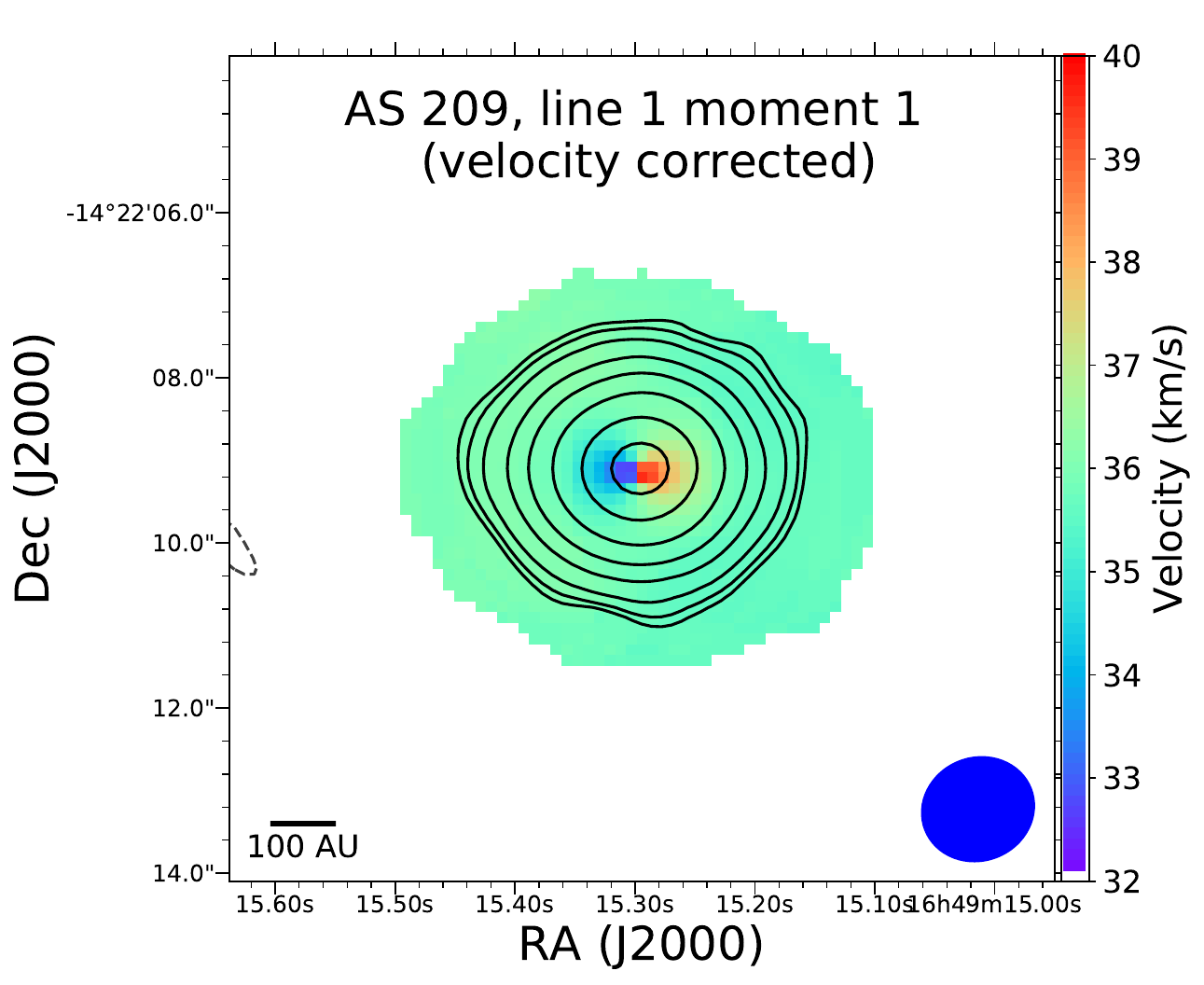}
    \caption{\added{Velocity (moment 1) map for hyperfine 1, made using data corrected for Keplerian rotation using {\tt \string gofish}. Contours represent total intensity (Stokes $I$) of 3, 10, 50, 100, 200, 325 and 500$\sigma$ levels.}}\label{mom1_shifted}
\end{figure}

However, unlike the case for the clouds studied by \citet{1996ApJ...456..217C}, the disk of AS 209 is rotating, which broadens the line emission and would impact any disk spatial averaging. To account for this, \deleted{and facilitate stacking} we used the package {\tt\string gofish} \citep{2019JOSS....4.1632T} to shift the velocities in the Stokes $I$ and Stokes $V$ cubes by the amounts appropriate to correct for Keplerian motion. Based on the disk's inclination angle, position angle, distance, and stellar mass, {\tt\string gofish} calculates the expected Keplerian velocity for each pixel in the image cube. This allows us to \replaced{stack}{shift} the line emission for each Stokes parameter onto a common centroid velocity range based on the disk's deprojected Keplerian rotation profile. \added{\citet{2018ApJ...868..113T} showed that AS 209's rotation profile differs from the Keplerian profile of a disk with the parameters listed in the introduction by at most $\pm$5\%. For the purposes of this paper, assuming a Keplerian rotation profile is sufficient to account for the large majority of the motion within the disk. Figure \ref{mom1_shifted} shows a moment 1 map created from an image cube whose pixels have been shifted along the velocity axis to correct for Keplerian motion using {\tt\string gofish}. This velocity correction process is limited by our angular resolution. Pixels within one beam of the disk's center will contain emission from both the redshifted and blueshifted sides of the disk, and will thus have a velocity closer to the central velocity of the line than they would in an image with infinite angular resolution. This leads to the overcorrection of the velocities near the center of the disk seen in Figure \ref{mom1_shifted}. We therefore exclude the region within a 0.7 arcsecond radius (half of the beam major axis) of the disk's center from our fits.}

\begin{figure*}[ht!]
\gridline{\fig{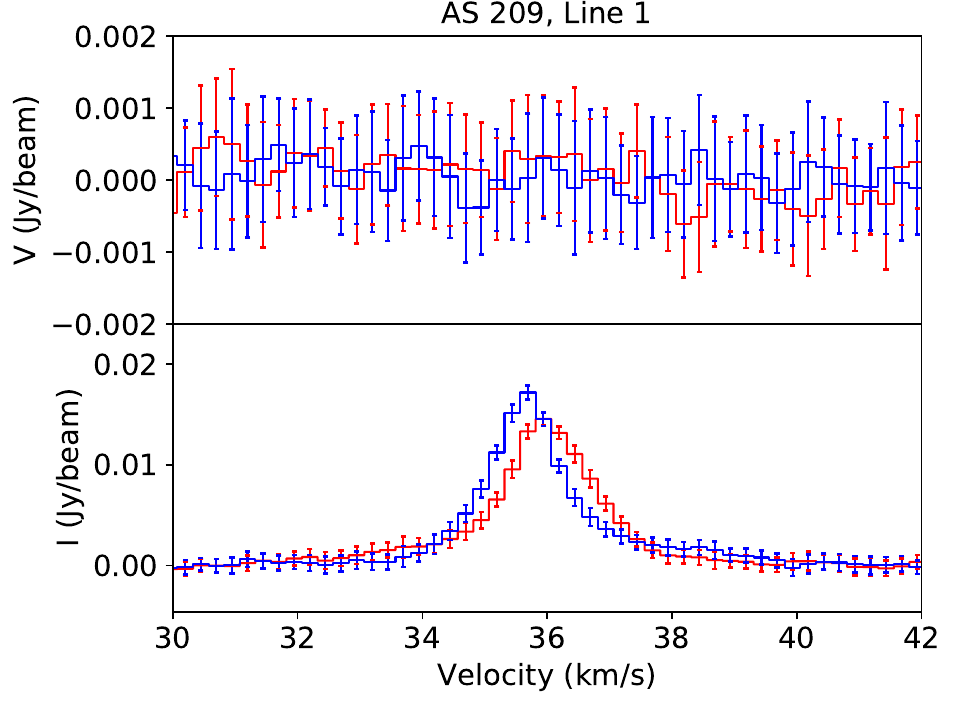}{0.3\textwidth}{(a)}
         \fig{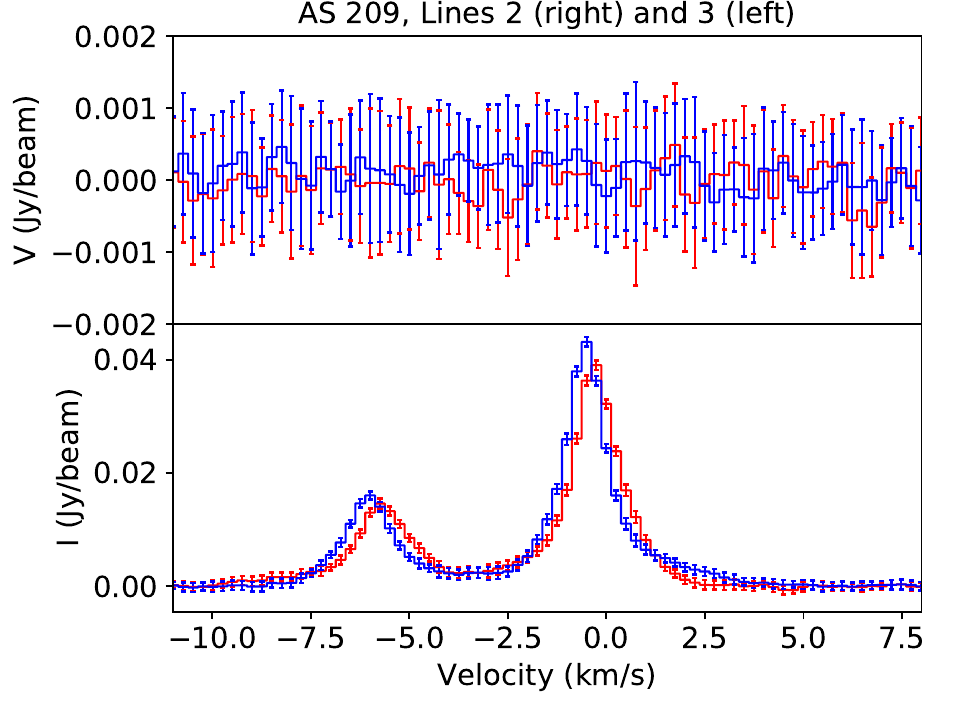}{0.3\textwidth}{(b)}
         \fig{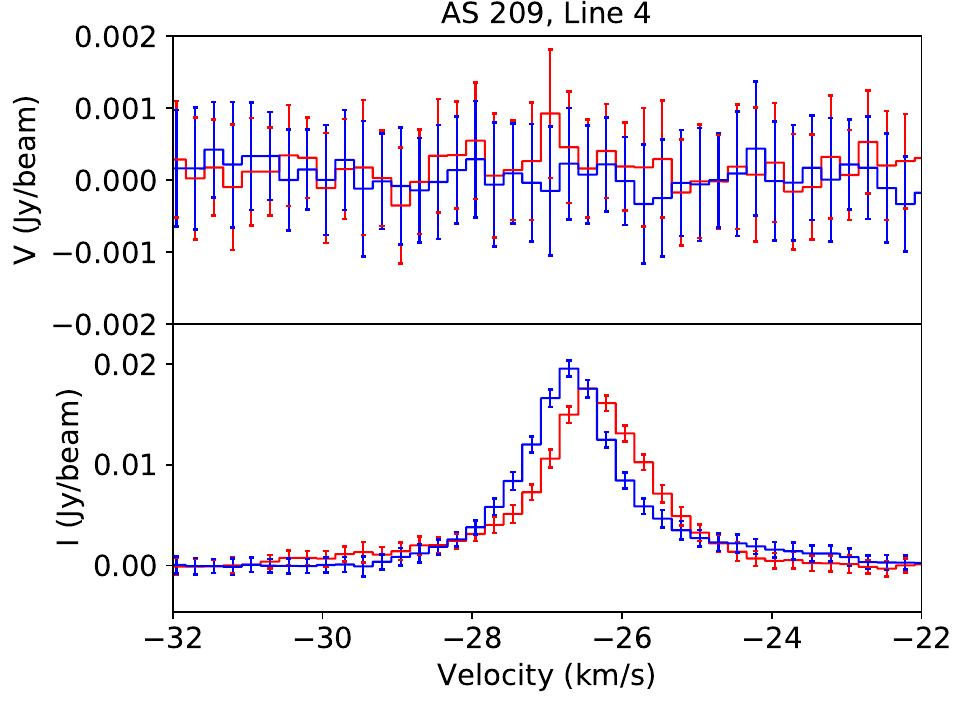}{0.3\textwidth}{(c)}}
\gridline{\fig{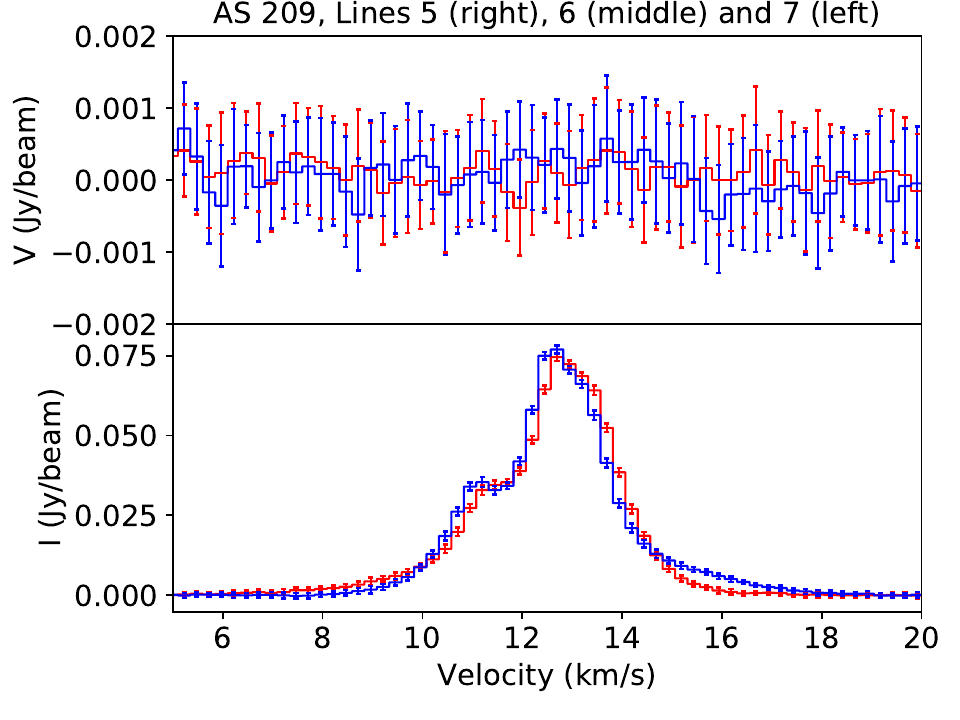}{0.3\textwidth}{(d)}
\fig{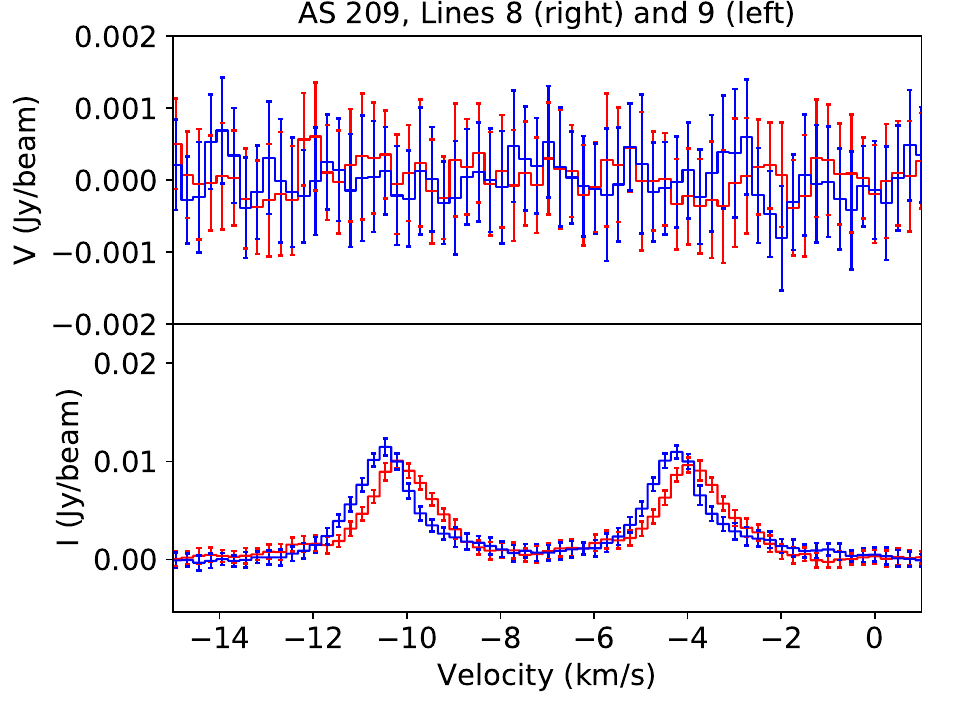}{0.3\textwidth}{(e)}}
          
\caption{\deleted{Stacked} Stokes $V$ and Stokes $I$ profiles created using {\tt\string gofish} for hyperfine components 1-9, averaged across the redshifted and blueshifted sides of AS 209. Velocities are LSRK velocities with respect to the velocity of hyperfine component 2. The red line represents the redshifted side of the disk, and the blue line represents the blueshifted side of the disk. \added{The sinusoidal fluctuations in the Stokes $V$ profiles are caused by the correlation of noise between velocity channels}}
\label{spectra}
\end{figure*}

Our angular resolution is sufficient to resolve the redshifted and blueshifted sides of the disk. This is advantageous for investigating any toroidal component of the magnetic field because we expect a toroidal magnetic field to have opposite signs on the red and blue sides of the disk. We used two different \deleted{stacking} methods to extract information about the magnetic field in the disk. First, we created average Stokes $I$ and average Stokes $V$ profiles for the redshifted and blueshifted sides of the disk and fit these data using the technique described earlier in this section. The Stokes $I$ and $V$ spectra for all of the hyperfine components \replaced{stacked}{created} using {\tt\string gofish} are shown in Figure \ref{spectra}. To calculate the uncertainties in each velocity bin, we calculated the per-channel rms in a region outside the disk from spectral line cubes whose velocities had been Kepler-corrected in {\tt\string gofish}. 

After using {\tt\string gofish} to account for line broadening due to Keplerian motion, lines 5, 6, and 7 were still blended. We therefore used only the lower-frequency portion of the Kepler-corrected line 5 data and the higher-frequency portion of the Kepler-corrected line 7 data in the \citet{1996ApJ...456..217C} fit. \replaced{The magnetic field strength that produced the best fit to these data was 2.0 $\pm$ 1.6 mG on the redshifted side of the disk and 0.9 $\pm$ 1.4 mG on the blueshifted side of the disk.}{The magnetic field strength that produced the best fit to these data was 1.9 $\pm$ 1.7 mG on the redshifted side of the disk and 1.0 $\pm$ 1.4 mG on the blueshifted side of the disk.} \added{The value of C$_1$ was -3.7$\times 10^{-3} \pm 1.1 \times 10^{-3}$ on the red side of the disk and $-4.1\times10^{-3}\pm 1.1\times 10^{-3}$ on the blue side. The value of C$_2$ were 1.8$\pm$2.3 Hz on the red side of the disk and -1.4$\pm$1.9 \added{Hz} on the blue side. The Stokes V profiles created by Zeeman splitting are dependent on the Zeeman splitting factors which are different for each line, making it unlikely that an instrumental effect like beam squint (which would affect all lines in the same way) could destroy a real Zeeman signal. The uncertainty of 0.8$^\circ$ on the disk's inclination angle adds an additional 0.07 mG to the 1$\sigma$ error bar on the net line-of-sight magnetic field strength on the redshifted side of the disk, and 0.03 mG on the blueshifted side of the disk. For the remainder of this paper, we will deal with the upper limits derived using an assumed inclination angle of 35.3$^\circ$.} Using the uncertainties on the magnetic field strengths as the 1$\sigma$ value, the limit on the net line-of-sight magnetic field strength is 5.0 mG on the redshifted side of the disk and 4.2 mG on the blueshifted side of the disk. \replaced{Given the disk's inclination angle of 35.3$^\circ$, this places 3$\sigma$ upper limits of 8.4 mG (red) and 8.0 mG (blue) on the net toroidal magnetic field and 5.9 mG (red) and 5.6 mG (blue) on the net vertical magnetic field.}{Given the disk's inclination angle of 35.3$^\circ$, this places 3$\sigma$ upper limits of 8.7 mG (red) and 7.3 mG (blue) on the net toroidal magnetic field and 6.3 mG (red) and 5.1 mG (blue) on the net vertical magnetic field.} The Stokes $I$ and $V$ \replaced{stacked}{averaged} spectra from the red and blue sides of each line are shown in Figure \ref{spectra}.

\begin{figure*}[ht!]
\gridline{\fig{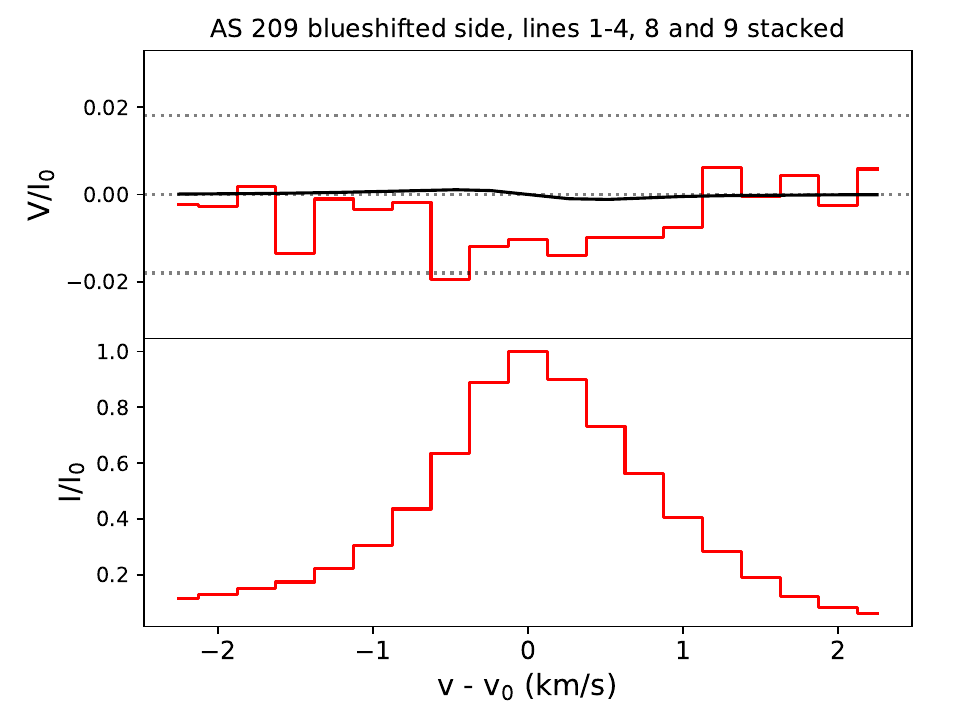}{0.45\textwidth}{(a)}
          \fig{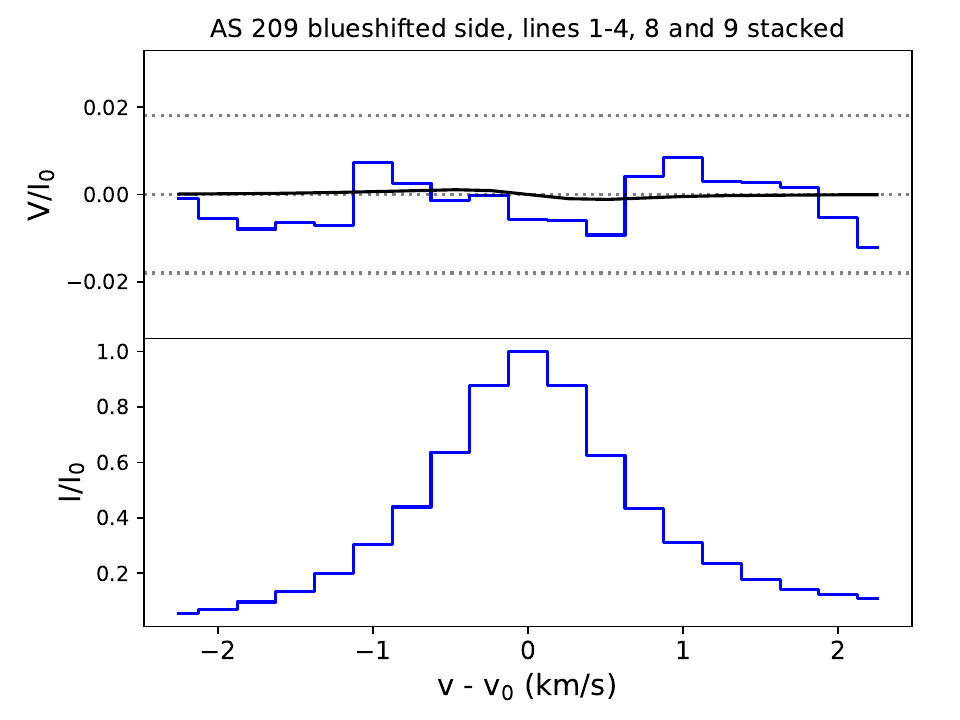}{0.45\textwidth}{(b)}
          }
\gridline{\fig{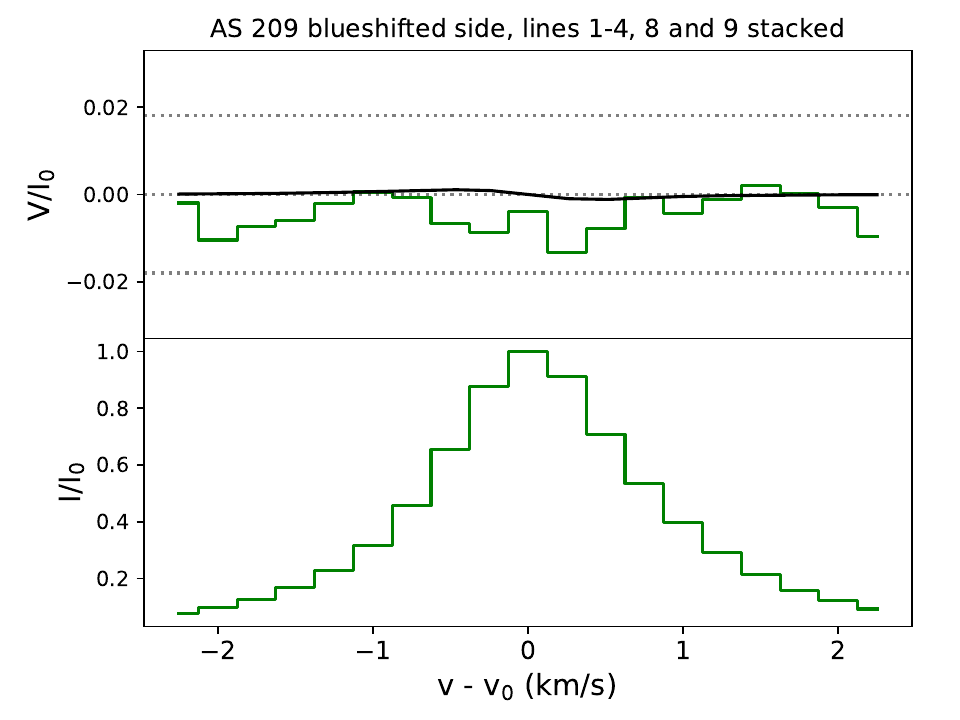}{0.45\textwidth}{(c)}}
\caption{Stacked Stokes $I$ and $V$ spectra using {\tt\string gofish} for the redshifted (a) and blueshifted (b) sides of the disk, and for the full disk (c). \added{The black lines represent the best fit of the equation $V = \frac{1}{2} Z B_{LOS}\frac{dI}{d\nu}$ to the data.} The dashed horizontal lines in the plots of $V/I_0$ represent $\pm$ 1.8\% of the $I/I_0$ peak.}
\label{stacked}
\end{figure*}

Second, we stacked \added{(i.e., summed)} the Stokes $I$ and Stokes $V$ spectra from both the red and blue sides of the disk for only the unblended lines. The Stokes $I$ spectra were scaled by their relative intensities, and the Stokes $V$ spectra were scaled by the relative Stokes $I$ intensities and the Zeeman splitting factors. If the disk field is toroidal, we would expect the Stokes $V$ spectrum from one side of the disk to be the same shape as the spectrum from the other side of the disk, but mirrored across the velocity axis. This is because the line-of-sight component of a toroidal field would have opposite signs on the redshifted and blueshifted sides of the disk, and $V(v)$ is proportional to the line-of-sight component of $B_z$, including the sign. Therefore, stacking $V(v)$ from the red side of the disk with $-V(v)$ from the blue side of the disk should increase the SNR of any Stokes $V$ emission that comes from a toroidal field. However, this method of stacking would destroy any Stokes $V$ signal from a vertical magnetic field, as we expect the direction of the vertical component of the field to remain the same across the disk. Nonetheless, stacking the lines in this way did not lead to a detection of circularly polarized emission. We fit the weighted sum of the line profiles $V_{tot} = \frac{1}{2I_0}B_{los} \sum_i Z_i \frac{dI_i}{d\nu_i} w_I w_Z$ to the stacked line data, where the $i$'s are lines 1, 2, 3, 4, 8, and 9; $I_0$ is the maximum Stokes $I$ intensity, and $w_I$ and $w_Z$ are weighting factors that account for the relative intensities of the lines and the relative strengths of their Zeeman splitting factors, respectively. The stacked spectra are shown in Figure \ref{stacked}. \explain{Corrected a typo in the script used to calculate best fits and updated values in text accordingly.} \replaced{The best fit values for $B_{los}$ from this stacking technique were -4.4 $\pm$ 7.7 mG for the redshifted side of the disk, 3.3 $\pm$ 5.2 mG for the blueshifted side of the disk, and for 1.0 $\pm$ 5.4 mG for the full disk.}{The best fit values for $B_{los}$ from this stacking technique were -4.2 $\pm$ 6.9 mG for the redshifted side of the disk, 2.0 $\pm$ 4.6 mG for the blueshifted side of the disk, and 0.4 $\pm$ 4.9 mG for the full disk.} Because this method does not account for circular polarization from instrumental effects, we report the results of using the first fitting technique described above as our final upper limits on the magnetic field strength.\\

\subsection{Continuum linear polarization}
\begin{figure*}[ht!]
\gridline{\fig{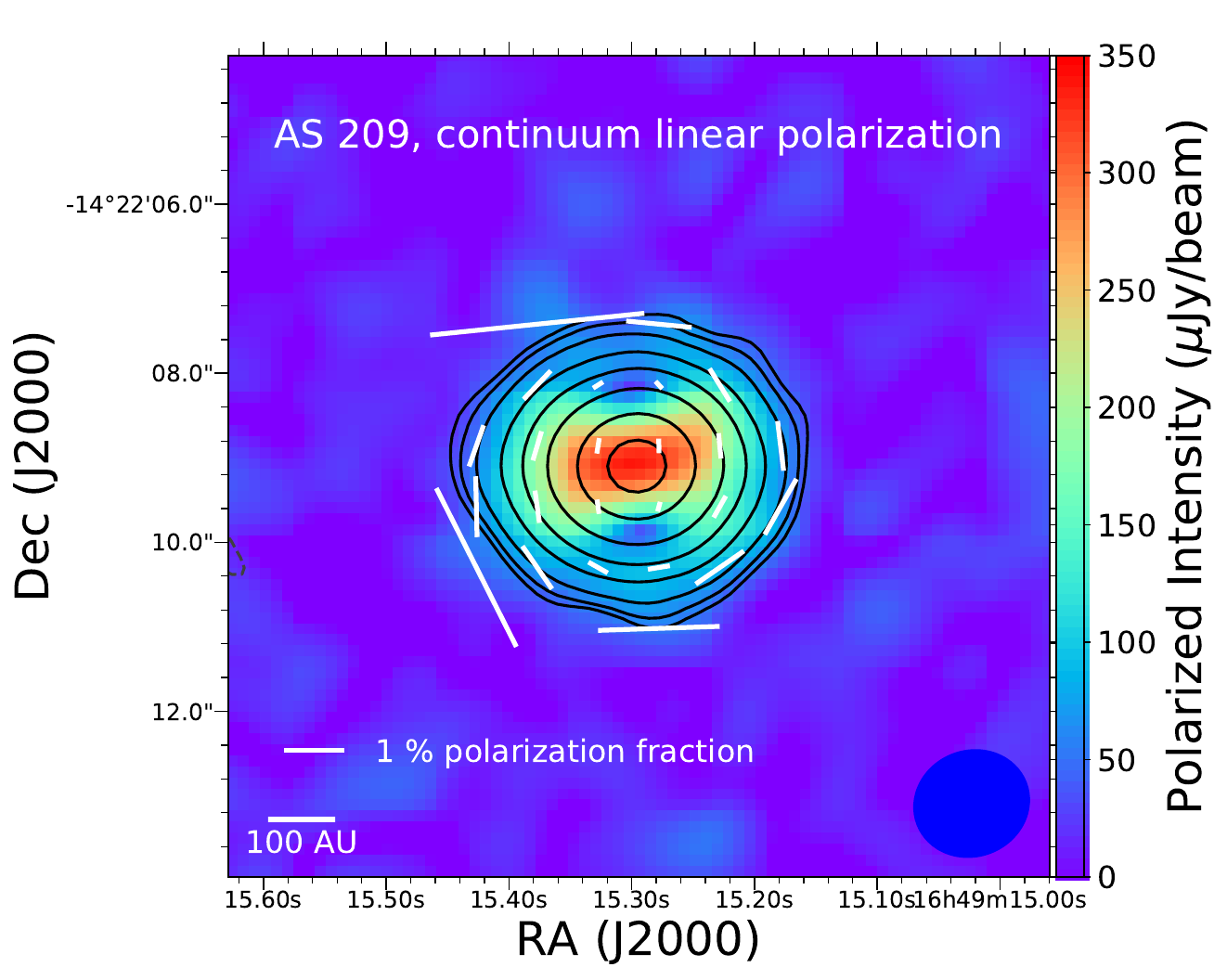}{0.45\textwidth}{(a)}
          \fig{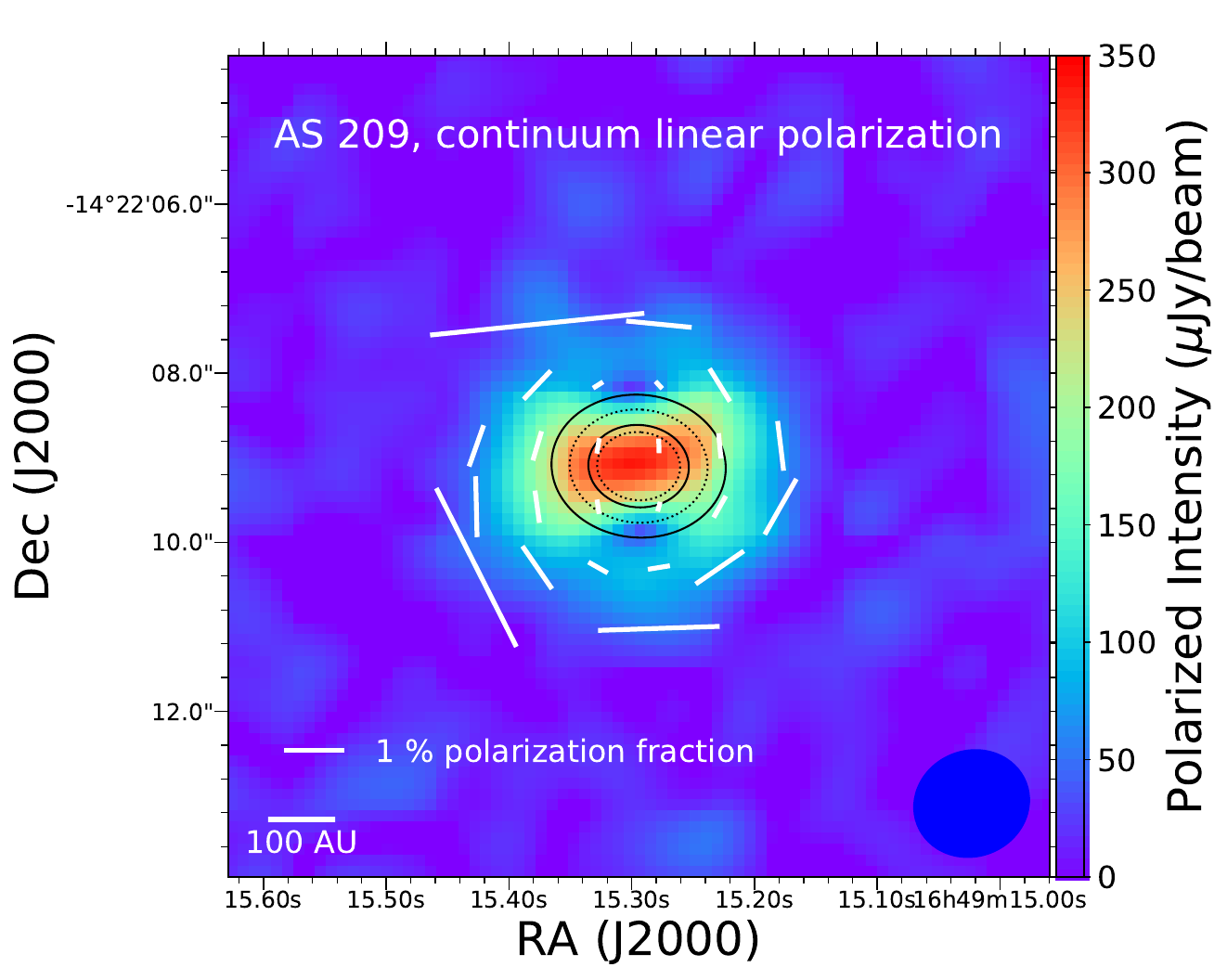}{0.45\textwidth}{(b)}
          }
\caption{1.3 mm continuum linear polarization in AS 209. The black contours in (a) represent total intensity (Stokes $I$) of 3, 10, 50, 100, 200, 325 and 500$\sigma$ levels. The solid contours in (b) represent the locations of rings, and the dashed contours represent the locations of gaps from \citep{2018A&A...610A..24F}. The colormap represents debiased polarized intensity with the scale on the right of each source. The length of the polarization vectors corresponds to the polarization fraction. The vectors are plotted with $\sim$ 5 segments per beam. Vectors are only plotted where the polarized intensity $P \geq 3 \sigma_P$ and $P/I \geq 0.1$.}
\label{contim}
\end{figure*}

The continuum polarization pattern in AS 209 at 1.3 mm is shown in Figure \ref{contim}.  Our observations are similar to those observed at 870 $\mu$m by \citet{2019ApJ...883...16M}, with the direction of polarization oriented parallel to the disk's minor axis in the inner part of the disk and oriented azimuthally in the outer part of the disk. This pattern closely matches the polarization from scattering in a disk with a similar inclination to AS 209's predicted by the model in \citet{2016MNRAS.456.2794Y}. 
We plan to explore the possible mechanisms behind this polarization pattern in a future paper.
   
\section{Discussion}


We did not detect circular polarization in any of the individual CN 2-1 lines or in the stacked lines, so we have calculated 3-$\sigma$ upper limits on the net toroidal and vertical magnetic field strengths. The minimum detectable degree of circular polarization with ALMA is 1.8\% of the peak Stokes $I$ flux according the ALMA Cycle 7 Technical Handbook. The Stokes $V$ flux does not reach this 1.8\% threshold in any channel of any individual hyperfine line or the stacked lines in our observations. Because the fitting technique described in Section 3.1 removes instrumental effects that could produce spurious Stokes $V$ signal, and because the Stokes $V$ spectrum is noise-like, the uncertainities on the magnetic field strengths calculated using the fitting technique can be used to calculate upper limits on the disk's magnetic field strength. The average rms values on the Stokes $V$ spectra shown in Figure \ref{spectra} were $\sim$0.9\% of the Stokes $I$ peak of the brightest line, which means that a Stokes $V$ signal $\geq$ 0.02 Jy beam$^{-1}$ would be detectable in the brightest line. Here, we discuss the implications of our field strength upper limits for the disk's mass accretion rate, as well as possible reasons for the non-detection. 

Our non-detection of Zeeman splitting in the CN 2-1 line allows us to put constraints on the mass accretion rate that the magnetic field can drive in the disk on the 10's of au scale. \replaced{Old discussion from ``If the magnetic stresses are responsible for driving the disk accretion through the magnetorotational instability (MRI), the total (toroidal and poloidal) magnetic field strength must be greater than" through ``In fact, the magnetic field has to be much weaker than 8 mG in order for the plasma-$\beta$ to have the values of $10^{3-5}$ that are often used in MHD disk simulations (see, e.g.,  \citet{suriano18}, and \citet{2014prpl.conf..411T} and references therein)."}{The magnetically driven accretion rate ${\dot M}_{\rm mag}$ is related to the magnetic stresses through, e.g., equation 18 of \citet{Wang_2019}. Making the simplifying assumptions that $\vert B_R \vert$, $\vert B_z \vert$, and $\vert B_\phi \vert$ are constant, and that $B_\phi$ has opposite signs above and below the midplane, this equation can be recast into a rough order-of-magnitude estimate of the magnetically driven accretion rate:

\begin{equation}
\begin{split}
    {\dot M}_{\rm mag} \approx \frac{2R}{\Omega}\vert B_z B_\phi\vert + \frac{2 h}{\Omega} \vert B_R B_\phi\vert \\
 =\frac{2R\vert B_z B_\phi\vert }{\Omega}\left(1+\frac{h}{R}\frac{\vert B_R\vert}{\vert B_z\vert}\right),
\end{split}
\label{eq:mdot1}
\end{equation}
where $B_R$, $B_z$ and $B_\phi$ are, respectively, the (cylindrically) radial, vertical and azimuthal component of the magnetic field, $h$ the disk scale height, and $\Omega=(GM_*/R^3)^{1/2}$ is the angular Keplerian speed at radius $R$. The first term on the right hand side of the equation is mass accretion driven by a magnetized disk wind and the second term is that from magnetic stresses internal to the disk. They are consistent with the estimates from \citealt[][see their equations 6 and 15 respectively]{2009ApJ...701..737B}. Since the disk is geometrically thin, with $h/R\ll 1$, the magnetic disk wind tends to remove angular momentum more efficiently than the internal magnetic stresses for comparable $B_z$ and $B_R$. In this case, we have 
\begin{equation}
\begin{split}
    {\dot M}_{\rm mag} \approx \frac{2R\vert B_z B_\phi\vert}{\Omega} = 2.1\times 10^{-6} \left(\frac{\vert B_z\vert}{6.2\ {\rm mG}}\right)\left(\frac{\vert B_\phi\vert}{8.7\ {\rm mG}}\right) \\
\left(\frac{M_*}{1.25\ M_\odot}\right)^{-1/2}\left(\frac{R}{50\ {\rm au}}\right)^{5/2}\ \frac{M_\odot}{\rm yr},
\end{split}
\label{eq:mdot2}
\end{equation}
where we have normalized the vertical and toroidal components of the magnetic field by their respective $3\sigma$ upper limits and the stellar mass $M_*$ by the value inferred in \citet{2018ApJ...868..113T}. \replaced{This}{The $\sim 10^{-6}$ $M_\odot$ yr$^-1$ value we estimate from our magnetic field strength upper limits} is to be compared with the mass accretion rate from the disk onto the central star of $10^{-7}$~$M_\odot$/yr estimated by \citet{2000ApJ...539..815J} based on the luminosity of ultraviolet lines (especially CIV). This accretion rate is on the high side for classical T Tauri stars and needs to be checked through independent methods.

It is possible that the absolute value of the disk magnetic field strength $\vert B\vert$ is substantially higher than our upper limits on the net line-of-sight magnetic field strength. This is particularly true in the case where the magnetic field in the disk is dominated by the toroidal component $B_\phi$ {\it and} $B_\phi$ reverses polarity across the disk midplane. Such a field reversal would be naturally produced if there is a net magnetic flux threading the disk (as is likely given that the disk forms out of magnetized dense cores that appear to have fairly regular magnetic fields as traced by dust polarization; for a recent review, see \citealt{2019FrASS...6....3H}). The differential rotation between the disk midplane and the  atmosphere and/or disk wind naturally twists the polodial field into a toroidal field that reverses direction around the midplane (for an illustration, see \citealt{suriano18}).
In this case, if the disk is not too optically thick, an individual line of sight will contain CN emission from above and below the disk midplane and thus sample gas with reversed toroidal magnetic fields and thus Zeeman signals of opposite sign. The optical depths of the bright rings of AS 209 are estimated to be 0.46 and 0.52 at ALMA Band 6 \citep{2018ApJ...869L..46D} and should be much lower in the gaps. It is likely that the bulk of the CN emitting materials both above and below the disk midplane contribute to the observed signals, which leads to a cancellation of the Zeeman signal. In this case, the Zeeman measurement can significantly underestimate the absolute strength of the magnetic field, as illustrated quantitatively by Mazzei et al. (2020, MNRAS, submitted).}

\section{Conclusions}
\replaced{We present the first Zeeman observations toward the circumstellar disk of AS 209, using 9 hyperfine components of CN 2-1 in ALMA Band 6. Although we easily detect the dust polarization of the disk, which is consistent with previous Band 7 observations, we do not detect any polarized emission in the CN lines. We used various stacking approaches, such as averaging the entire disk and averaging the blue/red sides of the disk, correcting for Keplerian rotation utilizing the code {\tt\string gofish}.  We present 3$\sigma$ upper limits based on the stacking technique described in \citet{1996ApJ...456..217C} because this technique allows us to remove circular polarization from instrumental effects from the Stokes $V$ spectra. In that case, we have calculated 3$\sigma$ upper limits on the net toroidal and vertical magnetic field strengths of $B_\phi < 8.4$ mG and $B_z < 5.9$ mG. While the change in direction of the toroidal magnetic field across the disk midplane may cause us to underestimate the absolute strength of the toroidal component of the magnetic field, estimates of the disk's plasma-$\beta$ indicate it is unlikely that the field strength is severely underestimated. Our constraints on the magnetic field strength provide an upper limit on the magnetically-driven mass accretion rate on the 50 au scale of order $10^{-8}$~M$_\odot yr^{-1}$ or smaller, which is an order of magnitude less than the mass accretion rate onto the star previously inferred for this object. }{We present the first Zeeman observations toward the circumstellar disk of AS 209, using 9 hyperfine components of CN 2-1 in ALMA Band 6. Although we easily detect the dust polarization of the disk, which is consistent with previous Band 7 observations, we do not detect any polarized emission in the CN lines. After correcting for the Keplerian rotation of the disk using {\tt\string gofish}, we used two approaches to derive upper limits on the magnetic field strengths: fitting the equation described in Equation \ref{eq:fit} to each hyperfine component as described in \citet{1996ApJ...456..217C} and fitting an equation with the form of the Zeeman splitting profile to the scaled sum of all of the un-blended components. We derived these limits for the redshifted and blueshifted sides of the disk, as well as the entire disk. We present 3$\sigma$ upper limits based on the stacking technique described in \citet{1996ApJ...456..217C} because this technique allows us to remove circular polarization from instrumental effects from the Stokes $V$ spectra. In that case, we have calculated 3$\sigma$ upper limits on the net toroidal and vertical magnetic field strengths of $B_\phi < 8.7$ mG and $B_z < 6.1$ mG. A change in \replaced{direction}{polarity} of the toroidal magnetic field across the disk midplane may cause us to underestimate the absolute strength of the toroidal component of the magnetic field, and therefore the true toroidal magnetic field strength could be $>$ 8.7 mG. Our constraints on the magnetic field strength provide an upper limit on the magnetically-driven mass accretion rate on the 50 au scale of order $10^{-6}$~M$_\odot yr^{-1}$ or smaller, which is consistent with the mass accretion rate onto the star previously inferred for this object.} 

\section{Acknowledgements}
This paper makes use of the following ALMA data: ALMA \#2018.1.01030.S. ALMA is a partnership of ESO (representing its member states), NSF (USA) and NINS (Japan), together with NRC (Canada), MOST and ASIAA (Taiwan), and KASI (Republic of Korea), in cooperation with the Republic of Chile. The Joint ALMA Observatory is operated by ESO, AUI/NRAO and NAOJ. The National Radio Astronomy Observatory is a facility of the National Science Foundation operated under cooperative agreement by Associated Universities, Inc. REH and LWL acknowledge support from NSF AST-1910364. ZYL is supported in part by NASA 80NSSC18K1095 and 80NSSC20K0533 and NSF AST-1716259 and AST-1910106. RMC is supported in part by NSF-1815987. We would like the North American ALMA Science Center data analysts for performing the initial data calibration and imaging. We would also like to thank Ryan Loomis and Mark Lacy for their support during our data reduction visit to the NAASC. \added{We greatly appreciate the comments from the editors and anonymous referees that significantly improved this \replaced{Letter}{paper}.}

\bibliographystyle{aasjournal}
\bibliography{citations}




\listofchanges
\end{document}